\documentclass[sigconf]{acmart}

\usepackage{amssymb}
\usepackage{amsmath}
\usepackage{listings}
\usepackage{graphicx}
\usepackage{multirow}
\usepackage{balance}
\usepackage{colortbl}
\AtBeginDocument{%
  }

\copyrightyear{2025}
\acmYear{2025}
\setcopyright{acmlicensed}\acmConference[ICMR '25]{Proceedings of the 2025 International Conference on Multimedia Retrieval}{June 30-July 3, 2025}{Chicago, IL, USA}
\acmBooktitle{Proceedings of the 2025 International Conference on Multimedia Retrieval (ICMR '25), June 30-July 3, 2025, Chicago, IL, USA}
\acmDOI{10.1145/3731715.3733296}
\acmISBN{979-8-4007-1877-9/2025/06}

\begin{document}

\title{Concept Drift Guided LayerNorm Tuning for Efficient Multimodal Metaphor Identification}


\author{Wenhao Qian}
\email{venh233.qian@gmail.com}
\affiliation{%
  \institution{Hefei University of Technology}
  \city{Hefei}
  \country{China}
}

\author{Zhenzhen Hu}
\authornote{Corresponding author.}
\email{huzhen.ice@gmail.com}
\affiliation{%
  \institution{Hefei University of Technology}
  \city{Hefei}
  \country{China}
}

\author{Zijie Song}
\email{zjsonghfut@gmail.com}
\affiliation{%
  \institution{Hefei University of Technology}
  \city{Hefei}
  \country{China}
}

\author{Jia Li}
\authornotemark[1]
\email{jiali@hfut.edu.cn}
\affiliation{%
  \institution{Hefei University of Technology}
  \city{Hefei}
  \country{China}
}

\renewcommand{\shortauthors}{Wenhao Qian, Zhenzhen Hu, Zijie Song, Jia Li}

\begin{abstract}
Metaphorical imagination, the ability to connect seemingly unrelated concepts, is fundamental to human cognition and communication. While understanding linguistic metaphors has advanced significantly, grasping multimodal metaphors, such as those found in internet memes, presents unique challenges due to their unconventional expressions and implied meanings. Existing methods for multimodal metaphor identification often struggle to bridge the gap between literal and figurative interpretations. Additionally, generative approaches that utilize large language models or text-to-image models, while promising, suffer from high computational costs.  This paper introduces \textbf{C}oncept \textbf{D}rift \textbf{G}uided \textbf{L}ayerNorm \textbf{T}uning (\textbf{CDGLT}), a novel and training-efficient framework for multimodal metaphor identification.  CDGLT incorporates two key innovations: (1) Concept Drift, a mechanism that leverages Spherical Linear Interpolation (SLERP) of cross-modal embeddings from a CLIP encoder to generate a new, divergent concept embedding. This drifted concept helps to alleviate the gap between literal features and the figurative task. (2) A prompt construction strategy, that adapts the method of feature extraction and fusion using pre-trained language models for the multimodal metaphor identification task. CDGLT achieves state-of-the-art performance on the MET-Meme benchmark while significantly reducing training costs compared to existing generative methods. Ablation studies demonstrate the effectiveness of both Concept Drift and our adapted LN Tuning approach. Our method represents a significant step towards efficient and accurate multimodal metaphor understanding. The code is available: \href{https://github.com/Qianvenh/CDGLT}{https://github.com/Qianvenh/CDGLT}.

\end{abstract}




\maketitle

\begin{figure*}
    \centering
    \includegraphics[width=\linewidth]{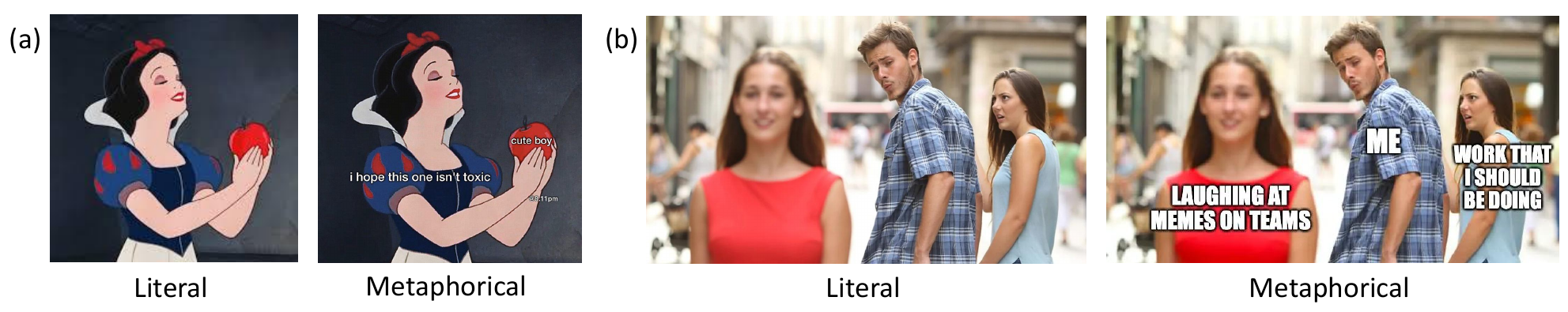}
    \caption{Concept Drift Phenomenon. Whether memes are metaphorical is closely related to the embedded text. (a) Before adding text: Snow White is about to take the apple. After adding text: The ``cute boy'' is likened to an apple that is about to be eaten, with the hope that new romantic interests aren't ``toxic''. (b) Before adding text: distracted boyfriend. After adding text: a joke about getting distracted from work responsibilities by looking at memes. It metaphorizes ``Work that I should be doing'' as neglected girlfriend and  ``Laughing at memes on teams'' as the distracting beauty.}
    \Description{}
    \label{fig:metaphorcase}
\end{figure*}

\section{Introduction}
Metaphorical imagination is the drawing of connections between seemingly unrelated domains to generate new meanings, insights, or creative expressions, often used to enhance understanding, communication, or artistic expression~\cite{lakoff2008metaphors}. It involves understanding one domain (often abstract or complex) in terms of another (typically more familiar or concrete), making it a fundamental feature of how we think and conceptualize the world~\cite{camp2006metaphor}. With the proliferation of multimodal content like posters and internet memes, metaphor has also extended to modalities such as visual forms~\cite{xu2022met, akula2023metaclue}.In recent years, although there has been significant progress in the identification and understanding of linguistic metaphors~\cite{stowe2021metaphor, li2023metaphor}, understanding multimodal or visual metaphors remains a challenge. This is because metaphors often involve unconventional expressions and implied meanings that go beyond their literal sense, and when these features are presented in visual form, the difficulties they pose are even greater.

The multimodal metaphorical meme dataset MET-Meme~\cite{xu2022met} has played a crucial role in advancing this field by catalyzing methodological innovations. The methods~\cite{he2024viemf, he2024sc, wang2024they} mainly focus on fine-grained feature alignment and fusion, resulting in suboptimal performance, as they often fail to fully leverage the characteristics of multimodal metaphors, overlooking the implied meanings and unconventional expressions that frequently appear in figurative tasks. On the other hand, methods~\cite{xu2024exploring, zhang2024camel, xu2024generating} that focus on generative knowledge expansion help bridge the gap between literal and figurative tasks by using the generative information from large language models (LLMs) or text-to-image models. Although using LLMs or text-to-image models has demonstrated significant advantages in multimodal metaphor identification tasks, they still face issues with high computational overhead and large GPU memory usage during training, even when these methods employ some general Parameter-Efficient Fine-Tuning (PEFT) techniques like LoRA~\cite{hu2021lora}. Recently, the novel methods~\cite{zhou2023one,yu2024aud} that only fine-tune the LayerNorm layers of language models for cross-modal feature extraction and fusion, have demonstrated outstanding efficiency: achieving good performance by fine-tuning less than 4\% of the total parameters. However, LayerNorm tuning of pretrained language models for feature extraction and fusion has remained unexplored in multimodal metaphor identification tasks due to their suboptimal performance when handling non-sequential data, such as images.

To address these issues, we propose a novel and training-efficient framework named \textbf{C}oncept \textbf{D}rift \textbf{G}uided \textbf{L}ayerNorm \textbf{T}uning (\textbf{CDGLT}) that introduces two key innovations for multimodal metaphor identification. First, we propose a lightweight and novel mechanism named ``Concept Drift'' to alleviate the gap between literal features and figurative tasks, based on an interesting phenomenon of the metaphorical meme that different texts embedded in the same image can change the metaphorical meaning of the meme (as intuitively demonstrated by the examples in Figure 1). Concept Drift constructs a new concept that has drifted from the original image features based on the OCR text features, serving as a divergent guide to assist in ``thinking outside the box''. Specifically, Concept Drift utilizes Spherical Linear Interpolation (SLERP)~\cite{shoemake1985animating} of two cross-modal embeddings from the CLIP~\cite{radford2021learning} encoder to produce an intermediate semantic embedding. Second, although LayerNorm Tuning (LN Tuning) has shown outstanding performance and efficiency in tuning language models for cross-modal feature extraction and fusion, it has primarily been applied to sequence information processing. In order to adapt it to the processing of non-sequential information, such as images, and then apply it to multimodal metaphor identification tasks, we devise a prompt construction strategy for LN Tuning that first fuses the features and then uses a frozen prompt to construct the sequence. This approach ensures effective feature fusion while fully utilizing the attention mechanism's ability to process sequences. Due to the small number of parameters required for training and no need for autoregressive iterative processing, the training of our model is highly efficient, solely requiring less than 5 minutes and under 5GB of GPU memory on a single RTX 4090.

The principal contributions of this work are threefold:
\begin{itemize}
\item We constructed a new concept embedding through SLERP as supplementary divergent information to help alleviate the gap between literal features and figurative tasks.
\item We leveraged our novel prompt construction strategy to adapt the feature extraction and fusion approach of LayerNorm Tuning the pretrained language model for the multimodal metaphor identification task, while also transferring their powerful sequence processing capabilities.
\item Our method achieved state-of-the-art performance on the MET-Meme benchmark. At the same time, through ablation experiments and analysis, we have demonstrated the effectiveness of the method we proposed.
\end{itemize}

\section{Related Work}

\subsection{Multimodal Metaphor Understanding}
Metaphor, a pervasive aspect of human language and communication, has been a subject of extensive research across various disciplines. Conceptual Metaphor Theory (CMT)~\cite{lakoff2008metaphors} posits that metaphors are not merely linguistic devices but reflect underlying conceptual mappings that shape our understanding of abstract concepts. Selectional Preference Violation (SPV)~\cite{wilks1975preferential} offers another perspective, highlighting how metaphors often involve a deviation from typical semantic expectations. The Metaphor Identification Procedure (MIP)~\cite{group2007mip} provides a structured approach for identifying metaphors in text, serving as a foundational methodology for metaphor analysis. In natural language processing (NLP), metaphor detection has gained increasing attention. Several computational models and approaches have been developed to automatically identify and interpret metaphors~\cite{stowe2021metaphor, ge2022explainable, mao2022metapro, mao2024metapro, li2023metaphor}.

The study of metaphors has expanded into the multimodal domain, exploring how this cognitive phenomenon manifests in various modalities, particularly in vision. Several datasets have been created to facilitate research in this area. MultiMET~\cite{zhang2021multimet} is a multimodal metaphor dataset containing text-image pairs with annotations for metaphor occurrence, domain relations, and sentiments. MetaCLUE~\cite{akula2023metaclue} is another dataset including four understanding tasks for multimodal metaphor. The MET-Meme~\cite{xu2022met} dataset provides a valuable resource for studying multimodal metaphors in the context of memes. MultiCMET~\citep{zhang2023multicmet} considers metaphors from different cultures, extending the work of MultiMET to Chinese. IRFL~\cite{yosef2023irfl} is a relevant dataset including metaphor and idiom data, which focuses on the multiple choices task. MemeCap~\cite{hwang2023memecap} includes annotations such as image captions, titles, and metaphorical captions to explore the automatic interpretation and understanding of multimodal metaphors. \cite{kalarani2024unveiling} introduces a task of video metaphor description, addressing the gap in Vision-Language (VL) models' ability to understand metaphors in video. Various models have been proposed for detecting and understanding multimodal metaphors. Early work focused on visual atypicality, which can be seen as a special kind of multimodal metaphor, such as \cite{guo2021detecting} models contextual compatibility to detect persuasive atypicality. The subsequent works primarily use MET-Meme as the benchmark, with early studies~\cite{he2024viemf, he2024sc, wang2024they, zheng2025multi} focusing on fine-grained multimodal feature fusion or alignment. Recently, there is a growing trend to leverage pre-trained generative models to expand the knowledge for metaphor understanding. For example, C4MMD~\citep{xu2024exploring} is a compact framework that uses a Chain-of-Thought (CoT) method to extract and integrate knowledge from Multimodal Large Language Models (MLLMs) into smaller models for multimodal metaphor detection. CAMEL~\cite{zhang2024camel} leverages the captions generated by multimodal language models as a bridge to capture the implicitly established metaphor alignment. And MMMC~\cite{xu2024generating} utilizes a text-conditioned generative adversarial network to generate visual characteristics based on the linguistic attributes of metaphorical concepts, thereby capturing more comprehensive visual associations. Lately, ImaRA~\cite{tian2025imara} leverages Retrieval-Augmented Generation to enhance LLMs’ capability to identify and interpret multimodal metaphors.

However, feature fusion-based methods often insufficiently consider the non-literal nature of metaphor tasks, while generative knowledge expansion methods tend to result in excessive computational and memory overhead during training. To address these issues, we propose a novel framework that leverages the feature space characteristics of pre-trained CLIP to construct a concept drift embedding as a divergent guide for figurative tasks (such as metaphor identification), while efficiently utilizing pre-trained GPT-2 for feature fusion and extraction.

\begin{figure*}[t]
    \centering
    \includegraphics[width=\linewidth]{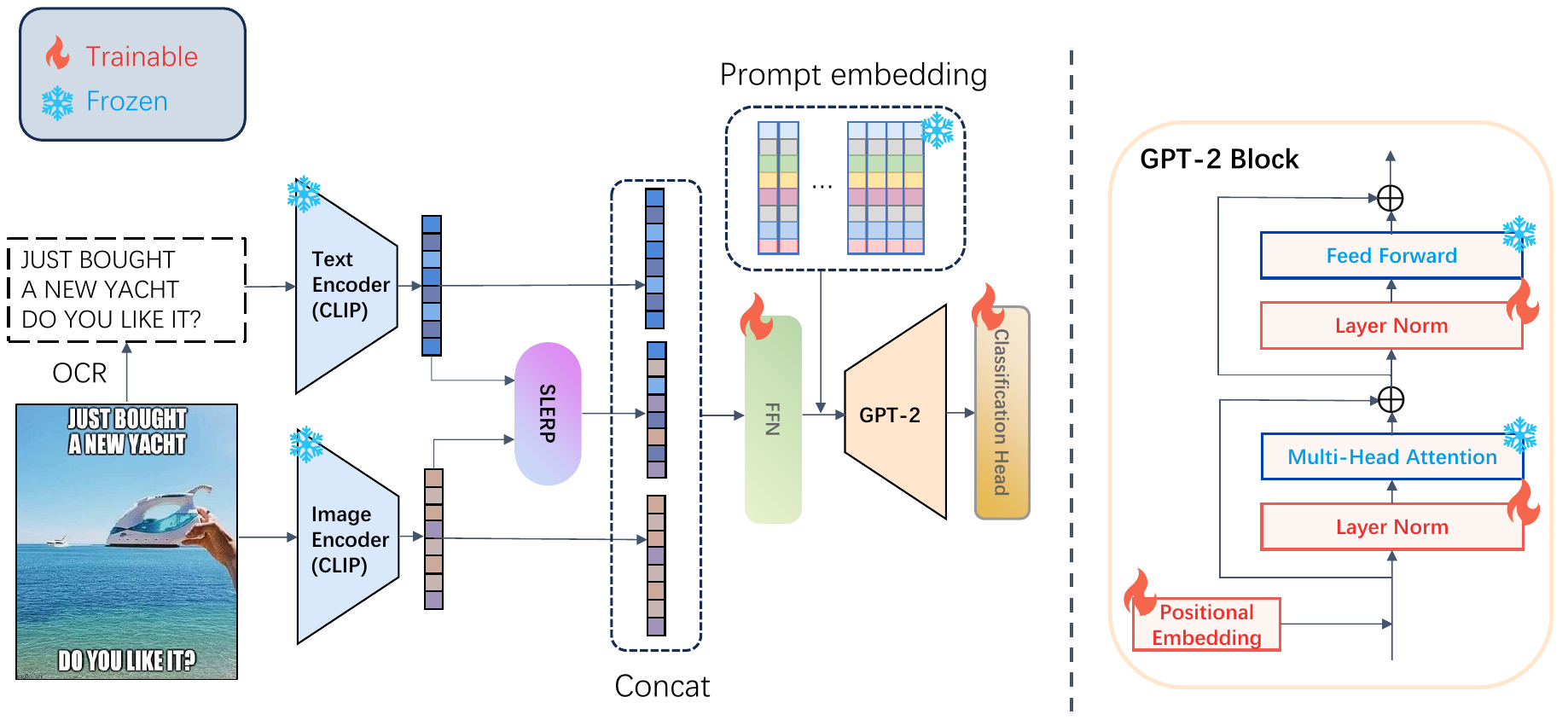}
    \caption{The architecture of CDGLT which is implemented with feature extraction, Concept Drift modeling, and LN tuning of GPT-2 using our novel prompt design.}
    \Description{}
    \label{fig:framework}
\end{figure*}

\subsection{LayerNorm Tuning}

Parameter-Efficient Fine-Tuning (PEFT) methods have been developed to adapt large pre-trained models to specific tasks with minimal computational overhead. Notable techniques include Adapter Tuning~\cite{houlsby2019parameter}, Low-Rank Adaptation (LoRA)~\cite{hu2021lora}, and Prompt Tuning~\cite{li2021prefix}. In addition to these methods, recent research has explored tuning LayerNorm parameters of pretrained Transformers~\cite{vaswani2017attention} as a means of efficient fine-tuning. And \cite{zhao2023tuning} introduces the efficient LN Tuning strategy for transforming Large Language Models (LLMs) into MultiModal Large Language Models (MLLMs). For zero-shot classification, \cite{li2024vision} introduces CLIPFit, a method that improves the performance of zero-shot CLIP by fine-tuning only specific bias terms and normalization layers. In addition, \cite{srirama2024hrp} only fine-tunes the LayerNorm Layer of ViT~\cite{dosovitskiy2021imageworth16x16words} to model the affordances. At the same time, there has been a trend of using LN Tuning pretrained language modals as a feature extraction and modality fusion strategy. For example, \cite{zhou2023one} presents a unified framework that fine-tunes the LayerNorm layer and the position embedding of GPT-2~\cite{radford2019language} for handling time series data. And \cite{yu2024aud} proposes a method that utilizes a pre-trained GPT-2 model for context-aware fusion of multimodal information to improve the accuracy of action unit detection. 

Our method also adopts the novel approach of LayerNorm Tuning with a pretrained language model as a feature extraction and fusion strategy. To leverage the powerful ability of attention in processing sequential information while adapting to the multimodal metaphor identification task, we propose a novel prompt construction method. Our experiments show that, in LayerNorm tuning the pretrained language model for the multimodal metaphor identification task, our method shows significant advantages compared to other prompt construction methods.

\section{Method}

The framework of CDGLT is depicted in Figure~\ref{fig:framework}. We employ the lightweight pretrained language model GPT-2~\cite{radford2019language} for meme image classification tasks, focusing on multimodal metaphor identification. In this section, we will detail the architecture of our model, which is composed of three parts: 1) Input processing and feature extraction; 2) Concept Drift and feature fusion; 3) LayerNorm tuning GPT-2.

\subsection{Input Processing and Feature Extraction}
Given an image $I$ and the extracted OCR text $T$, we utilize the frozen pretrained CLIP~\cite{radford2021learning} to obtain the image embedding $E^I \in \mathbb{R}^{N\times d_c}$ and text embedding $E^T \in \mathbb{R}^{N\times d_c}$: 
\begin{equation}
    E^I = \text{CLIP}_{\text{image}}(I)
\end{equation}
\begin{equation}
    E^T = \text{CLIP}_{\text{text}}(T)
\end{equation}
where $N$ represents the number of samples, and $d_c$ represents the common dimension shared by the two embeddings.

\subsection{Concept Drift and Feature Fusion}
The embedding spaces modeled by the CLIP image encoder and the text encoder are aligned. The angle or cosine similarity of embeddings from different modalities can be directly used to measure their semantic relationship. \cite{jang2025spherical} and \cite{han2024merlin} find that, such two 
CLIP embeddings like $E^I$ and $E^T$ can be used to construct an intermediate semantic embedding that is drifted away from the original embeddings through Spherical Linear Interpolation (SLERP)~\cite{shoemake1985animating}. We call the operation as Concept Drift whose output embedding can be seen as a new concept feature related to both the meme's image part and text part. Before applying the SLERP, the $E^I$ and $E^T$ should be L2-normalized to ensure that their magnitudes are the same (both equal to 1). After L2-normalizing $E^I$, the resulting vector is denoted as $\mathbf{v}$, while the L2-normalized version of $E^T$ is denoted as $\mathbf{w}$. The drifted embedding $E^S \in \mathbb{R}^{N\times d_c}$ can be constructed through the SLERP between $\mathbf{v}$ and $\mathbf{w}$ as:
\begin{equation}
    E^S = \frac{sin((1 - \alpha)\theta)}{sin(\theta)}\mathbf{v} + \frac{sin(\alpha\theta)}{sin(\theta)}\mathbf{w}
\end{equation}
where $\alpha \in [0, 1]$ is a hyperparameter, and $\theta$ is the angle between $E^I$ and $E^T$. Based on the need to deviate from the original image features, we follow the prior work~\cite{jang2025spherical} to set a text-weighted $\alpha$ as $0.8$ for composing the good drifted representation. The norm (or magnitude) of $E^S$ is the same as that of $\mathbf{v}$ and $\mathbf{w}$. $\theta$ is calculated as follows:
\begin{equation}
    \theta = \text{arccos}(\mathbf{v} \cdot \mathbf{w})
\end{equation}
After obtaining $E^S$, we fuse the three embeddings—$E^I$, $E^S$, and $E^T$—into a unified feature vector through concatenation and feedforward neural network (FFN) operations. The vector obtained after the concatenate operation is denoted as 
$E^{Meme} \in \mathbb{R}^{N\times 3d_c}$, and the vector obtained after the FFN is denoted as $F \in \mathbb{R}^{N\times d_g}$. This process is formalized as follows:
\begin{equation}
    E^{Meme} = \text{concat}(E^I, E^S, E^T)
\end{equation}
\begin{equation}
    F = \text{GELU}(E^{Meme}W_1+b_1)W_2 + b_2
\end{equation}
where $W_1 \in \mathbb{R}^{3d_c\times d_c}$ and $W_2 \in \mathbb{R}^{d_c\times d_g}$ are the learnable parameters, and $b_1$ and $b_2$ are the biases of the two linear layers of FFN. GELU is the activation function used. $d_g$ indicates the dimension of the hidden state in GPT-2.

\subsection{LayerNorm Tuning GPT-2}
In order to activate GPT-2's powerful feature extraction capabilities, which are mainly evident in sequence processing, we design an embedding sequence $P \in \mathbb{R}^{N\times (m+1)\times d_g}$ as prompt, which is composed of frozen Xavier initialization~\cite{glorot2010understanding} embeddings $E^x_i \in \mathbb{R}^{d_g}$ and the fused feature $F$ as:
\begin{equation}
    P = [E^x_0, E^x_1,..., E^x_m, F]
\end{equation}
where $m$ is a hyperparameter set as $10$ that represents the number of the frozen Xavier initialization embeddings. The assembled prompt $P$ serves as input to our GPT-2 model. Similar to~\citep{zhou2023one, yu2024aud}, our framework leverages the general feature extraction capability of the pre-trained GPT-2 model, which arises from large-scale text training. Fine-tuning only its Layer Normalization (LN) parameters and position embedding components allows us to both tailor the model for our specific multimodal metaphor identification task and maintain the generalization ability of the pretrained GPT-2. We utilize the last hidden state of GPT-2, $H^{(L)} \in \mathbb{R}^{N\times (m+1)\times d_g}$, to obtain the refine feature vector $F^\prime \in \mathbb{R}^{N\times d_g}$ :
\begin{equation}
    H^{(L)} = \text{GPT-2}_{LN\text{-}finetuned}(P)
\end{equation}
\begin{equation}
    F^\prime = \beta H^{(L)}_{-1} + (1-\beta) H^{(L)}_{-2} 
\end{equation}
where $L$ represents the number of blocks of the GPT-2 model, $H^{(L)}_{-1} \in \mathbb{R}^{N\times d_g}$ and $H^{(L)}_{-2} \in \mathbb{R}^{N\times d_g}$ respectively indicates the last and the second to last embeddings of $H^{(L)}$ and $\beta$ is a learnable weight. Finally, we feed $F^\prime$ into the classification head, which is a linear layer, to acquire the predicted probability $\hat{y}\in \mathbb{R}^{N\times k}$ :
\begin{equation}
    \hat{y} = \text{softmax}(F^\prime W_c+b_c)
\end{equation}
where $W_c \in \mathbb{R}^{d_g\times k}$ and $b_c$ are respectively the learnable parameter and the bias of the linear layer and $k$ is the number of the categories of the specific classification task. For example, in the Metaphor Identification task, $k = 2$.

\subsection{Loss Function}
To train our model for the task of 
the multimodal metaphor identification, we employ the Cross-Entropy (CE) loss, the loss function is as follows:
\begin{equation}
    \mathcal{L} = \frac{1}{N} \sum_{i=1}^{N} L_{CE}(y_i, \hat{y_i})
\end{equation}
where $N$ is the number of samples. $y$ represents the ground truth (GT) category.

\section{Experiments}
In this section, we conduct extensive experiments on various tasks on the benchmark focusing on multimodal metaphor identification. We first introduce the settings, including the dataset, evaluation metrics, and implementation details. Then we display the quantitative results with further ablation analysis.
\begin{table*}[ht]
\centering
\begin{tabular}{c|cccc|cccc}
\hline
\multirow{2}{*}{\centering Models} & \multicolumn{4}{c|}{W-F1$\uparrow$} & \multicolumn{4}{c}{Acc$\uparrow$} \\ \cline{2-9} 
            & SA  & OD  & ID  & MI  & SA  & OD  & ID  & MI  \\ \hline

$\text{MET}\_\text{add}$  \cite{xu2022met} &24.58 &67.28 &40.35 &81.41 &24.65 &68.39 &40.32 &81.33 \\
$\text{MET}\_\text{cat}$  \cite{xu2022met} &28.04 &66.70 &38.87 &82.54 &27.68 &67.25 &38.56 &82.39 \\
CAMEL-C \citep{zhang2024camel} &- &- &- &- &29.17 &72.24 &43.80 &85.06 \\
CAMEL-S \citep{zhang2024camel} &- &- &- &- &28.78 &72.36 &44.24 &85.57 \\
$\text{M}^\text{3}\text{F}\_\text{add}$ \citep{wang2024they} &31.89 &72.65 &43.11 &84.91 &30.47 &76.17 &44.40 &83.98 \\
$\text{M}^\text{3}\text{F}\_\text{cat}$ \citep{wang2024they} &31.85 &71.77 &44.33 &84.56 &29.82 &74.09 &44.10 &83.20 \\
MGMCF \cite{zheng2025multi} &35.98 &\textbf{77.92} &47.72 &88.04 &34.36 &\textbf{78.11} &47.92 & 87.51 \\
\cellcolor{yellow!30}$\text{CDGLT}_{Vanilla}$ (Ours) &\cellcolor{yellow!30}\underline{40.27} &\cellcolor{yellow!30}\underline{74.55} &\cellcolor{yellow!30}\textbf{51.06} &\cellcolor{yellow!30}\underline{90.81} &\cellcolor{yellow!30}\underline{40.38} &\cellcolor{yellow!30}\underline{76.95} &\cellcolor{yellow!30}\textbf{51.25} &\cellcolor{yellow!30}\underline{90.88} \\
\cellcolor{yellow!30}CDGLT (Ours) &\cellcolor{yellow!30}\textbf{42.28} &\cellcolor{yellow!30}72.92 &\cellcolor{yellow!30}\underline{49.30} &\cellcolor{yellow!30}\textbf{91.34} &\cellcolor{yellow!30}\textbf{41.00} &\cellcolor{yellow!30}74.35 &\cellcolor{yellow!30}\underline{49.38} &\cellcolor{yellow!30}\textbf{91.38} \\
 \hline
\end{tabular}
\caption{Main Results on the four tasks of MET-Meme: Sentiment Analysis (SA), Offensiveness Detection (OD), Intention Detection (ID) and Metaphor Identification (MI). The best scores are marked in bold, while the second best are underlined. We propose two variants of CDGLT based on the presence or absence of Concept Drift: CDGLT, which incorporate Concept Drift, and $\text{CDGLT}_{Vanilla}$, which does not.}
\label{tab:main-result}
\end{table*}

\begin{table}
\centering
\begin{tabular}{c|c|c}
\hline
\centering Models & Macro-F1$\uparrow$ & Acc$\uparrow$ \\ \cline{1-3} 

VIEMF \citep{he2024viemf} &83.92 &84.87 \\
SC-Net \citep{he2024sc} &86.85 &87.50 \\
C4MMD \citep{xu2024exploring} &79.05 &88.04 \\
ImaRA-7B \citep{tian2025imara} &83.45 &91.04 \\
\cellcolor{yellow!30}$\text{CDGLT}$ (Ours)
&\cellcolor{yellow!30}\textbf{88.84} &\cellcolor{yellow!30}\textbf{91.38} \\
 \hline
\end{tabular}
\caption{Comparison of Macro F1-score on the Metaphor Identification (MI). The best scores are marked in bold.}
\label{tab:main-result-macro}
\end{table}

\subsection{Dateset}
Considering both usability and relevance to multimodal metaphor, we follow previous research and select the \textbf{MET-Meme}~\cite{xu2022met} as the benchmark for our study. MET-Meme proposes four distinct classification tasks: sentiment analysis, intent detection, aggression detection, and metaphor identification including 4,000 English memes and 6,000 Chinese memes, each annotated with rich information about their metaphorical characteristics, including metaphor occurrence, sentiment, and offensiveness, etc. Although our approach primarily focuses on the Metaphor Identification task, our experiments found some interesting relationships in the results of CDGLT across the four tasks of MET-Meme. Additionally, reporting the experimental results of several tasks beyond Metaphor Identification also enables a more comprehensive comparison with prior work. And we only use the English part of MET-Meme in our work.

\subsection{Evaluation Metrics}
To assess the classification task performance, we report accuracy (Acc) and weighted F1-score (W-F1) as measurement indicators. Accuracy measures the ratio of correct predictions to the total number of test samples. The weighted F1 score offers a comprehensive evaluation by taking into account the support of each class. Consistent with previous research, the weighted F1 score is defined as the harmonic mean of the weighted averages of precision and the weighted averages of recall.
\begin{equation}
\text{Acc} = \frac{\sum_{i=1}^{L}TP_i}{N}
\end{equation}

\begin{equation}
Precision_i = \frac{(TP_i)}{(TP_i + FP_i)}
\end{equation}

\begin{equation}
Precision_{weighted} = \frac{\sum_{i=1}^{L} (Precision_i * w_i)}{|L|}
\end{equation}

\begin{equation}
Recall_i = \frac{(TP_i)}{(TP_i + FN_i)}
\end{equation}

\begin{equation}
Recall_{weighted} = \frac{\sum_{i=1}^{L} (Recall_i * w_i)}{|L|}
\end{equation}

\begin{equation}
\text{W-F1} = \frac{(2 * Precision_{weighted} * Recall_{weighted})}{(Precision_{weighted} + Recall_{weighted})}
\end{equation}

Where $L$ is denoted as the number of categories, N as the number of samples, $w_i$ as the weight (the proportion of samples) for the $i$th category, TP as true positives, TN as true negatives, FN as false negatives and FP as false positives.

\subsection{Implementation Details}
Our model was implemented in Pytorch~\cite{paszke2019pytorch} and all experiments were conducted on a single Nvidia RTX 4090 (24G) GPU. For the CLIP encoder, we used CLIP-ViT-L/14, where this alias refers to the model's parameter scale being large and the size of the image patch being 14. For the GPT-2 backbone, we used GPT2-base. All of the pretrained weights were provided by HuggingFace~\cite{thomas2019transformers}. We used the AdamW optimizer~\cite{loshchilov2017decoupled} with a fixed weight decay of 0.01 and the cosine learning rate scheduler~\cite{Loshchilov2016SGDR}. 
All models were trained for 200 epochs with early stopping based on the macro average F1-score on the validation set. For tasks with different numbers of classification categories, our architecture requires the replacement of different classification heads. To obtain the best combination of learning rate and batch size for different tasks, we conducted a hyperparameter search and used the accuracy of the checkpoint determined by the early stopping mechanism as the criterion for selecting the hyperparameter combinations. The learning rate was explored across three values: 1e-4, 5e-4, and 1e-3, while the batch size was evaluated at 96 and 128. We adopted the same dataset split as used in prior research~\cite{wang2024they} which randomly divided the dataset into training, validation, and test sets with a split of 60\%, 20\%, and 20\%, respectively. Our model demonstrates very efficient training, with GPU memory usage under 5GB. Using early stopping, the training time on the RTX 4090 takes less than 5 minutes.

\begin{table*}[ht]
\centering
\begin{tabular}{ccc|cccc|cccc}
\hline
\multirow{2}{*}{$E^I$}  &\multirow{2}{*}{$E^S$} &\multirow{2}{*}{$E^T$} & \multicolumn{4}{c|}{W-F1$\uparrow$} & \multicolumn{4}{c}{Acc$\uparrow$} \\ \cline{4-11} 
 & & & SA  & OD  & ID  & MI  & SA  & OD  & ID  & MI  \\ \hline

$\checkmark$ & & &40.29 &72.21 &50.16 &91.13 &39.12 &73.83 &50.38 &91.15 \\
&$\checkmark$ & &37.55 &73.16 &50.76 &85.75 &38.12 &74.87 &50.65 &85.94 \\
& &$\checkmark$ &35.05 &71.77 &49.49 &82.81 &35.62 &74.09 &49.61 &82.81 \\
$\checkmark$ &$\checkmark$ & &38.61 &73.59 &49.52 &91.44 &39.32 &74.22 &49.74 &91.38 \\
&$\checkmark$ &$\checkmark$ &35.71 &71.23 &50.26 &82.64 &35.88 &73.62 &50.26 &82.81 \\
$\checkmark$ & &$\checkmark$ &40.27 &74.55 &51.06 &90.81 &40.38 &76.95 &51.25 &90.88 \\
$\checkmark$ &$\checkmark$ &$\checkmark$ &42.28 &72.92 &49.30 &91.34 &41.00 &74.35 &49.38 &91.38 \\
\hline
\end{tabular}
\caption{The impact of different combinations of embeddings concatenated into $E^{Meme}$ on four tasks.}
\label{tab:ablation-1}
\end{table*}

\subsection{Main Results}
In this section, we compare our method with several existing multimodal metaphor detection methods, including MET-Meme~\cite{xu2022met}, CAMEL~\citep{zhang2024camel}, $\text{M}^\text{3}\text{F}$~\citep{wang2024they}, MGMCF~\cite{zheng2025multi}, VIEMF~\citep{he2024viemf}, SC-Net~\citep{he2024sc}, C4MMD~\citep{xu2024exploring}, and ImaRA~\citep{tian2025imara}. Among these methods, VIEMF, SC-Net, C4MMD and ImaRA focus exclusively on the metaphor identification (MI) task and report macro-averaged F1 scores. The remaining methods are evaluated on four tasks in the MET-Meme dataset: Sentiment Analysis (SA), Offensiveness Detection (OD), Intention Detection (ID) and Metaphor Identification (MI). At the same time, MET-Meme, $\text{M}^\text{3}\text{F}$ and CAMEL have all proposed two model variants, and we have also included these variants in our comparison. As shown in Table~\ref{tab:main-result} and Table~\ref{tab:main-result-macro}, we report the performance results of the methods to be compared and our model settings. One of our settings is without SLERP called $\text{CDGLT}_{Vanilla}$ and the other with SLERP just called CDGLT. Although our approach is primarily focused on the multimodal metaphor identification task, we found that our architecture, with specifically training on other tasks, also achieved performance superior to or on par with the state-of-the-art. It is noted that CDGLT achieves the highest accuracy and Weighted F1 score on MI and SA tasks. On the other hand, $\text{CDGLT}_{Vanilla}$ demonstrates the complementary situations, achieving the better performance on the other two tasks: ID and OD. This suggests that incorporating Concept Drift embedding enhances the model's ability to capture non-literal understanding tasks, such as MI. In contrast, tasks like ID and OD appear to benefit more from the vanilla model without SLERP, suggesting that these tasks demand more straightforward information and less divergence. This implies that the benefits of SLERP are not universally applicable and are instead highly dependent on the specific task. Our ablation analysis in section~\ref{sec:Concept-Drift-Analysis} will further clarify this situation. In addition, it can be observed that even with the use of Concept Drift, the CDGLT model achieves excellent performance in nearly all tasks, outperforming previous methods except in the accuracy of the OD task. Our framework does not utilize the generative information of LLM, surpassing methods like C4MMD, CAMEL and ImaRA, which use pre-trained language models to generate explanations of metaphors or implicit information as supplementary knowledge.

\subsection{Further Analysis}

In this subsection, extensive ablation studies have been conducted to verify the effectiveness of different parts. At the same time, we will introduce some interesting findings.

\subsubsection{Concept Drift Analysis.}
\label{sec:Concept-Drift-Analysis}
In this section, we conducted a series of ablation experiments on the composition of $E^{Meme}$ and observed some interesting phenomena. As shown in Table~\ref{tab:ablation-1}, the results compare the input of the FFN adapter, $E^{Meme}$, with different settings. First, comparing ``$E^I$ only'' versus ``$E^T$ only'', the results of the two experiments show only minor differences in the OD and ID tasks, while there are significant differences in the SA and MI tasks. Combined with other experiments, it is clear that the image modality is indispensable for improving performance across almost all tasks. This aligns with our intuition, as memes themselves use images as carriers of all information, including text. However, interestingly, for ID and OD tasks, just the OCR text from memes, although not the highest, still performs quite well. At the same time, this also suggests that, compared to the other two tasks, there may be more compatible commonalities between SA and MI, and the same applies to OD and ID.

The two experiments, ``$E^{S}$ only'' and ``concat($E^I$, $E^T$)'', reveal that on the SA and MI tasks, the performance of ``concat($E^I$, $E^T$)'' closely aligns with that of ``$E^I$ only''. Meanwhile, the performance of ``$E^S$ only'' just falls between that of ``$E^I$ only'' and ``$E^T$ only''. This may suggest that the concatenated feature vector aggregates information from both embeddings. However, the $E^S$ obtained through the SLERP operation seems to represent an intermediate state transitioning from $E^I$ to drift towards $E^T$. 

\begin{figure}
    \centering
    \includegraphics[width=\linewidth]{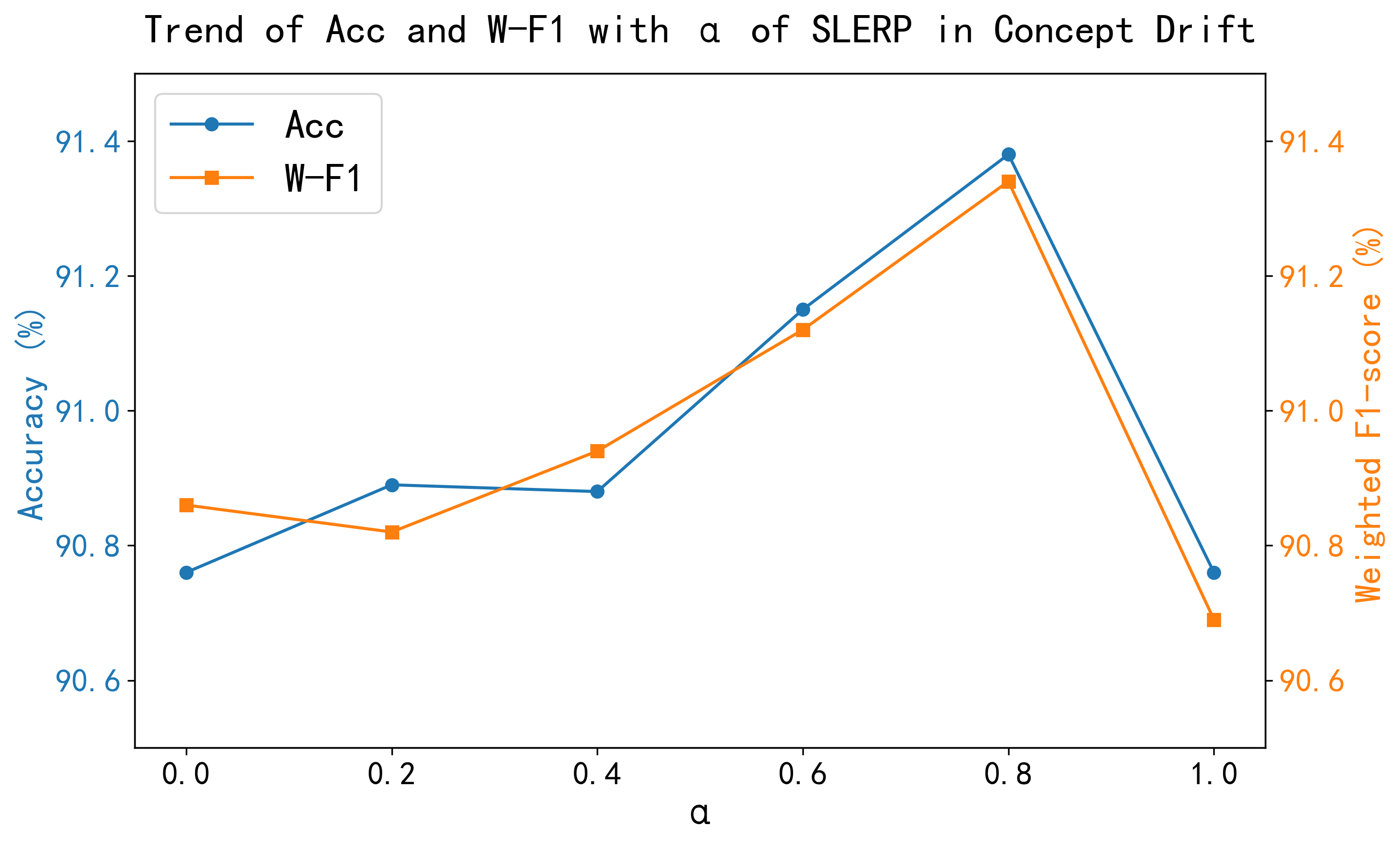}
    \caption{Trend of accuracy and weighted F1-score with $\alpha$ of SLERP in Concept Drift on Metaphor Identification (MI) task.}
    \Description{}
    \label{fig:linechart}
\end{figure}

However, for the ID and OD tasks, there is a phenomenon where performing concatenation between any two embeddings of $E^I$, $E^T$, and $E^S$, results in improved performance on the ID or OD tasks compared to using $E^I$ or $E^T$ alone. Specifically, ``concat($E^I$, $E^S$)'' improves performance on the OD task compared to using ``$E^I$ only'', while ``concat($E^T$, $E^S$)'' enhances performance on the ID task compared to ``$E^T$ only''. Finally, ``concat($E^I$, $E^T$)'' leads to improvements in both the ID and OD tasks. Similar to ``concat($E^I$, $E^T$)'', using ``$E^S$ only'' also leads to improvements in both the ID and OD tasks compared to using ``$E^I$ only'' and ``$E^T$ only''. Additionally, from the results of the group ``concat($E^I$, $E^S$, $E^T$)'' and ``concat($E^I$, $E^T$)'', it can be observed that while adding $E^S$ improves the performance of SA and MI tasks, it also leads to a decline in the performance of ID and OD tasks. Furthermore, in the ID task, ``concat($E^I$, $E^S$, $E^T$)'' performs worse than ``concat($E^T$, $E^S$)'', and in the OD task, it performs worse than ``concat($E^I$, $E^S$)''. Therefore, concatenating any two embeddings with a third embedding will lead to a decrease in performance on both the OD and ID tasks. This could mean that the diverging information might be redundant for both OD and ID tasks, or even become noise that interferes with performance.

Comparison of the two experiments ``concat($E^I$, $E^S$)'' and ``concat($E^I$, $E^T$)'', it shows that although replacing $E^T$ with $E^S$ leads to poorer performance on other tasks, it results in an improvement on the MI task. Additionally, by comparing ``concat($E^I$, $E^S$)'' with ``$E^I$ only'' or ``concat($E^I$, $E^S$, $E^T$)'' with 'concat($E^I$, $E^T$)', it is found that adding $E^S$ also contributes positively to the performance on the MI task. At the same time, this may suggest that divergent guidance, such as Concept Drift, indeed contributes to figurative tasks.

\begin{figure}
    \centering
    \includegraphics[width=\linewidth]{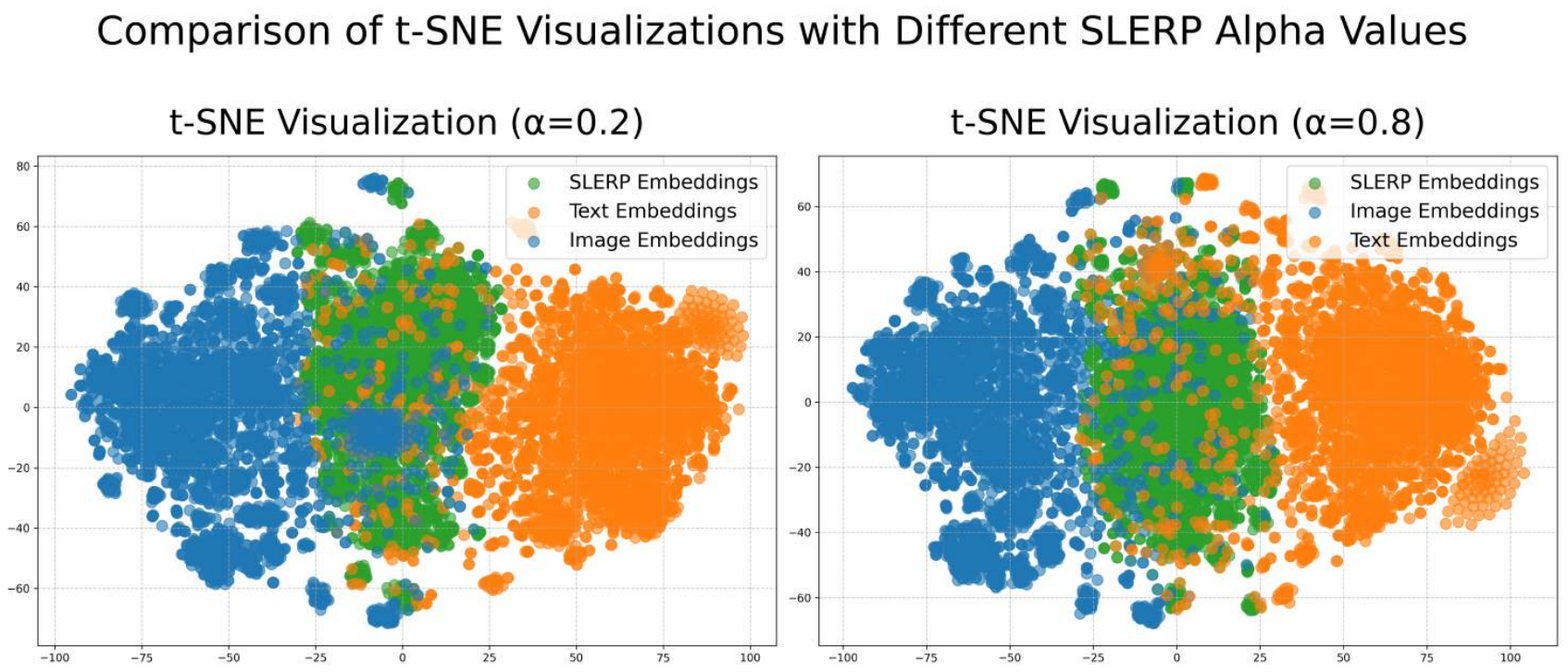}
    \caption{t-SNE visualizations of CLIP image, text and SLERP embeddings from MET-Meme training set. The left image shows the results for alpha = 0.2, while the right image shows the results for alpha = 0.8. The colors of the points represent different embedding types: image embeddings (blue), text embeddings (orange), and SLERP embeddings (green).}
    \Description{}
    \label{fig:tsne}
\end{figure}

Finally, for Slerp, we vary the balancing value $\alpha$ with six different settings, 0.0, 0.2, 0.4, 0.6, 0.8 and 1.0 for $E^S$ of concat($E^I$, $E^S$, $E^T$) on the MI task, and display the results in Figure~\ref{fig:linechart}. From these results, we observe that increasing the weight assigned to text embedding enhances both accuracy and the weighted F1-score performance, peaking at $\alpha = 0.8$. However, when the image has no effect ($\alpha = 1.0$), there is a significant drop in performance, bringing it back to the level of $\alpha = 0.0$. Moreover, the performance of both ($\alpha = 0.0$ and $\alpha = 1.0$) is comparable to that of ``concat($E_I$, $E_T$)''. To further investigate whether SLERP can cause the obtained embeddings to drift away from image embeddings, we conducted t-SNE visualizations of CLIP image, text and SLERP embeddings (reducing from 768 dimensions to 2) from MET-Meme training set using different SLERP alpha values, 0.2 and 0.8, as shown in Figure~\ref{fig:tsne}. It can be observed that the three embeddings exhibit clear clustering patterns. Compared to alpha = 0.8, when alpha = 0.2, there is more overlap between SLERP embeddings (green) and image embeddings (blue). When alpha is set to 0.8, the overlap between SLERP embeddings (green) and text embeddings (orange) increases, while the overlap with image embeddings (blue) decreases, indicating a trend of drifting away from image features. Therefore, it can be concluded that the supplementary new concept embedding, when text-weighted or, in other words, drifting from the image, is more advantageous for the metaphor identification task.

\begin{table}
\centering
\begin{tabular}{l|c|c|c}
\hline
Prompt &length & W-F1$\uparrow$ & Acc$\uparrow$ \\
\hline
Vision Tokens &257 &87.39 &87.50 \\
{$E^I$} &1 &90.45 &90.49 \\
\hline
$F$ &1 &90.74 &90.75 \\
\hspace{0.25cm}+Instruction A: &14+1 &90.60 &90.62 \\
\hspace{0.25cm}+Trainalbe Vectors: &14+1 &90.69 &90.76 \\
\hspace{0.25cm}+Frozen Vectors: &14+1 &91.03  &91.02 \\ \hline
\hspace{0.25cm}+Instruction B: &10+1 &91.35 &91.41 \\
\hspace{0.25cm}+Trainalbe Vectors: &10+1 &90.71 &90.75 \\
\hspace{0.25cm}+Frozen Vectors: &10+1 &91.34  &91.38 \\ \hline
\hspace{0.25cm}+Instruction C: &5+1 &91.04 &91,02 \\
\hspace{0.25cm}+Trainalbe Vectors: &5+1 &90.74 &90.75 \\
\hspace{0.25cm}+Frozen Vectors: &5+1 &91.04  &91.02 \\
\hline
\end{tabular}
\caption{Prompt Ablations on Metaphor Identification (MI) task. ``Visual Tokens'' represent all the CLIP vision token features. $E^I$ represents the ``CLS'' token of CLIP vision encoder feature. Both are processed by a FFN adapter before being fed into GPT-2. F is the feature from concatenating $E^T$, $E^S$, and $E^I$ into an embedding three times the hidden dimension, then processed by the FFN adapter. The ``length'' column represents the number of embeddings of the sequence. Instruction A: ``Providing you with the memes, now identify the metaphorical ones:''; Instruction B: ``Metaphor Detection (literal | metaphor):''; Instruction C: ``Metaphor Detection:''. Trainable and Frozen Vectors refer to prompts constructed using trainable and frozen Xavier initialization vectors, respectively. }
\label{tab:ablation-2}
\end{table}

\subsubsection{Prompt Ablations.} In this section, we conduct a series of ablation studies on the prompt construction method we proposed for LN Tuning. Experimental results on the MI task is shown in Table~\ref{tab:ablation-2} demonstrating that, compared to other strategies, our method has significant advantages on the multimodal metaphor identification task based on meme data. The method of LN Tuning with a pretrained language model is commonly used for feature extraction or feature fusion of sequence information. Therefore, the most intuitive and common approach is to directly use all image patches (vision tokens) as the input to the language model. However, we found that by first fusing the multimodal global features (``CLS'' tokens) using an FFN adapter and then directly using the fused features $F \in \mathbb{R}^{N \times 1 \times d_g}$ as input, we achieved better performance, even though the input sequence length was 1 at that point. Since the attention mechanism exhibits stronger capabilities when processing sequences, we additionally construct an embedding sequence as a prompt based on $F$. We tried three strategies: the first one uses the word embedding layer of GPT2 to transform fixed instruction words into an embedding sequence; the second one uses Xavier initialization to create some learnable vectors; the third one freezes the Xavier-initialized vectors. In each method, $F$ is placed at the end of the sequence. For the first strategy, we tried three instructions: ``Providing you with the memes, now identify the metaphorical ones:'', ``Metaphor Detection (literal | metaphor):'' and ``Metaphor Detection:''. After tokenization, they consist of 14 tokens, 10 tokens and 5 tokens, respectively. For both ``Trainable vectors'' and ``Frozen vectors'', we also used sequence lengths of 14, 10 and 5, respectively. The experimental results show that ``frozen vectors'' perform significantly better than ``trainable vectors''. Comparing the experimental results of the Frozen Xavier initialization vectors method under different length settings may indicate that using prompts that are too long or too short could potentially lead to a decline in performance. However, we intuitively believe that when using word instructions as prompts, the performance will be significantly affected by the semantic content, and this impact is difficult to capture. At the same time, we noticed that in our experiments, the method of frozen vectors and word instructions performed similarly. To ensure the generality of the method, our framework ultimately chose to construct prompts by adopting frozen vectors.

\begin{table}
\centering
\begin{tabular}{c|c|c}
\hline
Backbone scale & W-F1$\uparrow$ & Acc$\uparrow$ \\
\hline
 GPT-2 Base (6)  &91.39  &91.25 \\
 GPT-2 Base &91.34 &91.38 \\
 GPT-2 Large &90.92 &90.88 \\
\hline
\end{tabular}
\caption{The impact of different parameter sizes of GPT-2. ``GPT-2 Base (6)'' refers to the use of only the first 6 blocks of the GPT-2 Base model.}
\label{tab:ablation-3}
\end{table}

\begin{table}
\centering
\begin{tabular}{c|c|c|c}
\hline
SLERP &Encoder & W-F1$\uparrow$ & Acc$\uparrow$ \\
\hline
\multirow{4}{*}{\centering $\times$} &BERT \& ResNet &86.98 &87.11 \\
&BERT \& ViT &85.56 &85.68 \\
&CLIP-ViT-B/32 &90.94 &91.00 \\
&CLIP-ViT-L/14 &90.81  &90.88 \\
\hline
\multirow{2}{*}{\centering  $\checkmark$} &CLIP-ViT-B/32 &90.96 &91.02 \\
&CLIP-ViT-L/14 &91.34  &91.38 \\
 \hline
\end{tabular}
\caption{The effect of different types of image encoder and text encoder.}
\label{tab:ablation-4}
\end{table}

\subsubsection{Pretrained Model Ablations.} In this section, we conducted some ablation studies on pretrained models, including examining the impact of different parameter sizes of GPT-2 and the effects of different types of encoders. The performance of different sizes of GPT-2 on the MI task is presented in Table~\ref{tab:ablation-3}. GPT-2 Base has 12 blocks, and the ``GPT-2 Base (6)'' in the table refers to using only the first 6 blocks of GPT-2 Base. Based on the results presented, we find that both larger and smaller parameter sizes of GPT-2 may lead to suboptimal performance. In addition, we attempted to replace CLIP-ViT-L/14 with another version, CLIP-ViT-B/32. We also conducted some experiments by replacing CLIP with BERT~\cite{devlin2018bert}, ResNet~\cite{he2015deepresiduallearningimage}, and ViT~\cite{dosovitskiy2021imageworth16x16words}. Specifically, the version of BERT used was BERT-base, ResNet was ResNet-50, and ViT was ViT-B-16. These image encoders or text encoders do not perform multimodal unified feature space modeling, so they do not meet the assumptions for using SLERP. However, as shown in Table~\ref{tab:ablation-4}, even in our non-SLERP framework, using the CLIP model yields better performance. As analyzed earlier in the section~\ref{sec:Concept-Drift-Analysis}, the performance on the MI task largely depends on the image encoder's performance. For tasks like multimodal metaphor identification, which extend linguistic phenomena to images, it is not surprising that visual models such as CLIP, which are trained with natural language supervision, can achieve better performance.

\section{Conclusion}

In this work, we introduced CDGLT, a novel and training-efficient framework designed for multimodal metaphor identification in internet memes. CDGLT effectively addresses key challenges in this domain by combining a unique Concept Drift Modeling approach with parameter-efficient LayerNorm Tuning, enhanced by a novel prompt construction strategy. Our primary contributions encompass three key aspects: (1) the introduction of a Concept Drift mechanism, leveraging SLERP to generate divergent semantic embeddings, which effectively bridges the gap between literal visual features and figurative meanings inherent in metaphorical expressions; (2) a prompt construction strategy that adapts the feature extraction approach of LayerNorm Tuning the pretrained language model for the multimodal metaphor identification task; and (3) the demonstration of state-of-the-art performance on the widely-used MET-Meme benchmark, validating the effectiveness of our approach. Through comprehensive ablation studies, we further confirmed the individual contributions of both Concept Drift and our tailored prompt for LayerNorm Tuning. This work significantly advances multimodal metaphor research by harmonizing computational efficiency, enhanced interpretability, and superior performance.

\begin{acks}
This work was supported by the NSFC NO. 62172138 and No.62202139. This work was also partially supported by the Fundamental Research Funds for the Central Universities NO. JZ2024HGTG0310.
\end{acks}

% \\vfill\eject

\bibliographystyle{ACM-Reference-Format}
\balance
\bibliography{base}

\end{document}